\documentclass[preprint,prd]{revtex4}%
\usepackage{amsfonts}
\usepackage{amsmath}
\usepackage{amssymb}
\usepackage{graphicx}%
\begin{document}
\title{Dynamics of redshift/blueshift during free fall under the Schwarzschild horizon}
\author{H.V.Ovcharenko}
\affiliation{Department of Physics, V.N.Karazin Kharkov National University, 61022 Kharkov, Ukraine}
\affiliation{Institute of Theoretical Physics, Faculty of Mathematics and Physics, Charles
University, Prague, V Holesovickach 2, 180 00 Praha 8, Czech Republic}
\author{O.B.Zaslavskii}
\affiliation{Department of Physics and Technology, Kharkov V.N. Karazin National
University, 4 Svoboda Square, Kharkov 61022, Ukraine}

\begin{abstract}
We consider a free-falling observer who crosses the event horizon in the
Schwarzschild background. In the course of this fall, he/she can receive
signals from an object (like a star surface) that emits radiation. We study
how the frequency received by an observer changes depending on the proper time
on his/her trajectory. The scenarios are classified depending on whether the
frequency is infinite, finite or zero near the singularity and the horizon.
This depends crucially on the angular momenta of an observer and a photon. In
this work we consider also emission process, and, as we show, conditions of
emission strongly influence parameters of a photon, and thus received
frequency. As one of our main results, we present numerical calculations
showing evolution of the received frequency during the process of diving into
a black hole, depending on parameters of an observer and emitter. We also
analyze how a falling observer will see a night sky as he/she approaches the
singularity. We show that there appear several blind zones, which were not
analyzed previously.

\end{abstract}
\maketitle

\section{Introduction}

\bigskip

\qquad One of the most beatiful and intriguing consequences of general
relativiy is the existence of black holes. They are also encountered in many
other gravity theories. As a consequences, a spacetime has nontrivial causal
structure, so the inner part is unaccessible for an external observer (as long
as quantum effects are neglected). However, as this region does exist and can
be, at least in principle, studied by local observers (if they are brave
enough to dive into a black hole), it is of interest to consider in detail
what can be found by such observers. This would also help us to understand,
what is going on in realistic scenarios of gravitational collapse, to which
the picture of an eternal black hole can serve as a reasonable idealization.

There is rather vast literature on properties of spacetime inside a black
\ hole (see, e.g. \cite{hamin} and references therein). In recent decade, this
issue revealed one more interesting aspect. It is connected with high energy
collisions. Mainly, such collisions were considered near the horizon
\cite{ban}. Meanwhile, it is of especial interest what happens near the
singularity and whether or not such collisions near it can change the fate of
a black hole due to backreaction. In this context, the frequency of a photon
measured by an observer can be thought of as a collision between a massive
particle (observer) and a photon. Study of such collisions for massive
particles showed here that the behavior of the energy in the center of mass
frame depends strongly on relation between dynamic characterstics of particles
(especially, their angular momenta \cite{flow}, \cite{3}).

In a previous work \cite{gr20} general features of such a redshift/blueshift
were described, with emphasis made on the vicinity of the horizon and
singularity. In the present work we perform next step and relate such a
frequency shift to scenario of evolution in the course of which an observer
moves through a part of spacetime inside a black hole. The main goal of this
investigation is to study how frequency of radiation registered by a
freely-falling observer under the horizon of a Schwarzschild black hole
changes as the observer approaches the singularity. Also, we discuss the view
of night sky saw by a free-falling observer and how its angular momentum
affects it.

The paper is organized as follows. In Sec. \ref{scheme} we describe, in
general, how we calculate the frequency of radiation received by an observer,
if it is known in the point of emission. In Sec. \ref{gen} we write the form
of the metric under discussion outside and inside the horizon. In the latter
case, we give explicitly the tetrad attached to an observer. In Sec. \ref{eqs}
we write equations of motion of massive and massless particles under the
horizon. In Seq. \ref{em} we relate the frequency of a photon in different
frames. In Sec. \ref{par} we relate the frequency and angular momentum of a
photon conserved along its trajectory with similar local quantities in the
frame comoving with an emitter that is crucial for the properties of radiation
near the singularity. In Sec. \ref{fr} we \ give the general expression for
the frequency of a photon that generalizes \cite{gr20} to the case when an
receiver and a photon move in different planes. In Sec. \ref{sky} we describe
how an observer under the horizon sees night sky. In Sec. \ref{star} we
discuss the trajectory of a photon using the Novikov coordinate frame. In Sec.
\ref{kr} we do the same in the Kruskal coordinates. In Sec. \ref{cond} we
discuss the kinematic conditions necessary for absorption of a photon to
occur. In Sec. \ref{res} we give the results of numeric calculation with their
subsequent discussion: how the frequency measured by a free-falling observer
changes along its trajectory. In Sec. \ref{sum} we give summary of results.
In Sec. \ref{ap} we give Appendix in which we suggest straightforward
derivation of the formula of aberration under the horizon.

\section{General scheme of the approach \label{scheme}}

In this investigation, we consider different origins of radiation: either it
comes from stationary observers outside the horizon (with prescribed
parameters of photons) or it is emitted by a falling body (for example, by a
surface of a collapsing star). Thus, we have three objects that are involved
in this process: an emitter, photons emitted by it, and an observer. To find
the frequency of the photon registered by an observer, we have to solve
several intermediate problems.

\begin{itemize}
\item Solve the equations of motion for the emitter and observer.

\item Relate the frequency of the photon, emitted in the frame comoving with
the emitter, to the frequency in a stationary frame (or its counterpart under
the horizon).

\item Investigate the evolution of the frequency with respect to stationary
observers along the path of a photon during its propagation under the horizon.

\item Find the point of collision between the observer and the photon.

\item Determine the frequency registered by the observer.
\end{itemize}

This scheme allows us to solve the initial problem of the relation between the
frequency of the registered photon and the properties of the emitter.

\section{General relations \label{gen}}

Let us consider the Schwarzschild metric.
\begin{equation}
ds^{2}=-fdt^{2}+\dfrac{dr^{2}}{f}+r^{2}d\Omega^{2}\text{,}%
\end{equation}
where $f=1-\frac{r_{+}}{r}$, $r_{+}$ being the horizon radius. We are
specifically interested in the region beyond the horizon, and for this
purpose, we will use the following coordinates (see \cite{nov61}): $t=y,~r=-T$:%

\begin{equation}
ds^{2}=-\dfrac{dT^{2}}{g}+gdy^{2}+T^{2}d\Omega^{2}\text{,} \label{met2}%
\end{equation}
$g=-f\geq0$.

In what follows, we will use the system of fiducial observers who remain at
rest with respect to the metric (\ref{met2}). Therefore, for these observers,
$y=const$, $\phi=const$. To them, we attach the tetrad%
\begin{gather}
e_{(0)\mu}=\left(  -\frac{1}{\sqrt{g}},0,0,0\right)  \text{,\ \ \ }e_{(1)\mu
}=\left(  0,\sqrt{g},0,0\right)  \text{,}\label{tetr}\\
e_{(2)\mu}=\left(  0,0,\left\vert T\right\vert ,0\right)  ,\text{\ \ }%
e_{(3)\mu}=\left(  0,0,0,\left\vert T\right\vert \sin\theta\right)  .
\end{gather}

It is worth noting that such observers are geodesic and have no counterpart in
the region outside the horizon (see \cite{dor} for details).

\section{Equations of particle motion \label{eqs}}

First of all, we will consider the motion of a massive particle in the
interior of a black hole. This analysis can be applied to both the emitter and
the observer. We denote the mass of a particle $m$ and its four-velocity
$u^{\mu}$. Since a black hole is spherically symmetric, we can always choose a
plane in which the motion occurs. As usual, it is convenient to choose it to
be $\theta=\frac{\pi}{2}$, so $u^{\theta}=0$. As the metric (\ref{met2})\ is
invariant under translations along the $y$ axis and rotations in $\varphi$
direction, the corresponding quantities $u_{y}=p$ and $u_{\varphi}%
=\mathcal{L}$ are conserved, where $p$ and $\mathcal{L}$ are the momentum in
$y$-direction and angular momentum per unit mass, respectively. If a particle
with the specific energy $\varepsilon$ enters the inner region from the
outside, the meanings of energy and momentum are interchange in accordance
with the interchange of temporal and spatial coordinates, so $\varepsilon=p$.

The normalization condition gives us
\begin{equation}
g_{\mu\nu}u^{\mu}u^{\nu}=-1=\dfrac{p^{2}}{g}-\dfrac{(u^{T})^{2}}{g}%
+\dfrac{\mathcal{L}^{2}}{T^{2}}\text{,}%
\end{equation}
whence
\begin{equation}
u^{T}=Z,\text{\ \ }\mathrm{w}\mathrm{h}\mathrm{e}\mathrm{r}\mathrm{e}%
~Z=\sqrt{p^{2}+g\left(  1+\dfrac{\mathcal{L}^{2}}{T^{2}}\right)  }\text{,}
\label{Z_expr}%
\end{equation}
so in coordinates $(T,y,\phi,\theta)$
\begin{equation}
u^{\mu}=\left(  Z,\dfrac{p}{g},\dfrac{\mathcal{L}}{T^{2}},0\right)  .
\label{u}%
\end{equation}

Tetrad components $(a=0,1,2,3)$
\begin{equation}
u^{(a)}=u^{\mu}e_{\mu}^{(a)}=\left(  \frac{Z}{\sqrt{g}},\frac{p}{\sqrt{g}%
},\dfrac{\mathcal{L}}{\left\vert T\right\vert },0\right)  . \label{tet}%
\end{equation}

In a similar way, we can consider the motion of a massless particle (photon).
Let it have the four-vector $k^{\mu}$. As the vector $k^{\mu}$ is null,
$g_{\mu\nu}k^{\mu}k^{\nu}=0$, whence%

\begin{equation}
k^{\mu}=\left(  z,\dfrac{q}{g},\dfrac{l}{T^{2}}0\right)  , \label{km}%
\end{equation}
where
\begin{equation}
z=\sqrt{q^{2}+g\dfrac{l^{2}}{T^{2}}}\text{,}%
\end{equation}
$q$ and $l$ are conserved.

\section{Emission of photon \label{em}}

Now let us relate the frequency of the emitted photon with respect to the
stationary observer and the emitter. To do this, we will relate the
characteristics of the emitter's motion in the original coordinate frame
(\ref{met2}) and the comoving frame. In the comoving frame, the corresponding
components of the four-velocity of the emitter should be $u^{(a)}\text{{}%
}^{\prime\prime}=(1,0,0,0)$. We can make transition to this frame in several steps.

\begin{itemize}
\item Take the velocity components in the tetrad frame (\ref{tet}).
\end{itemize}

\begin{itemize}
\item Rotate:
\begin{gather}
u^{(y)}\text{{}}^{\prime}=u^{(y)}\cos\phi+u^{(\varphi)}\sin\phi\text{,}\\
u^{(\varphi)}\text{{}}^{\prime}=u^{(\varphi)}\cos\phi-u^{(y)}\sin\phi
\text{,}\\
u^{(T)}\text{{}}^{\prime}=u^{(T)}\text{.}%
\end{gather}
We adjust the rotation angle to have $u^{(\varphi)}${}$^{\prime}=0$. This
gives us
\begin{equation}
\tan\phi=\dfrac{u^{(\varphi)}}{u^{(y)}}=\dfrac{\mathcal{L}\sqrt{g}%
}{p\left\vert T\right\vert }, \label{phi}%
\end{equation}
$u^{(y)}\text{{}}^{\prime}=\sqrt{(u^{(y)})^{2}+(u^{(\varphi)})^{2}}=\dfrac
{1}{\sqrt{g}}\sqrt{Z^{2}-g}$.\newline
\end{itemize}

\begin{itemize}
\item Boost:
\begin{gather}
u^{(T)}\text{{}}^{\prime\prime}=\cosh\eta u^{(T)}\text{{}}^{\prime}-\sinh\eta
u^{(y)}\text{{}}^{\prime}\text{,}\\
u^{(y)}\text{{}}^{\prime\prime}=\cosh\eta u^{(y)}\text{{}}^{\prime}-\sinh\eta
u^{(T)}\text{{}}^{\prime}=0\text{,}\\
u^{(\phi)}\text{{}}^{\prime\prime}=u^{(\phi)}{}^{\prime}=0.
\end{gather}
This gives us
\begin{equation}
\tanh\eta=\dfrac{u^{(y)}\text{{}}^{\prime}}{u^{(T)}\text{{}}^{\prime}}%
=\dfrac{\sqrt{(u^{(y)})^{2}+(u^{(\varphi)})^{2}}}{u^{(T)}}, \label{eta}%
\end{equation}
\newline As 4-velocity normalization condition gives $u^{(T)}\text{{}}%
^{2}-(u^{(y)}\text{{}}^{\prime})^{2}=1$, we find $\cosh\eta=u^{(T)}\text{{}%
}^{\prime}=\dfrac{Z}{\sqrt{g}}$, $\sinh\eta=u^{(y)}\text{{}}^{\prime}%
=\dfrac{1}{\sqrt{g}}\sqrt{Z^{2}-g}.$

\qquad Now, we can find similar transformations for a photon. First, we assume
that the photon propagates in the same plane as the emitter. In the reference
frame comoving with the emitter, the photon has the wavevector $k^{(a)}%
\text{{}}^{\prime\prime}=k(1,\cos\psi,\sin\psi,0)$, where $k>0$. Going back to
the frame (\ref{met2}), we have:
\begin{gather}
k^{(T)}=k(\cosh\eta+\sinh\eta\cos\psi),\label{k_fr}\\
k^{(y)}=k[(\sinh\eta+\cosh\eta\cos\psi)\cos\phi-\sin\psi\sin\phi],\\
k^{(\phi)}=k[\sin\psi\cos\phi+(\sinh\eta+\cosh\eta\cos\psi)\sin\phi]\text{.}%
\end{gather}
Using (\ref{phi}) and (\ref{eta}), one finds
\begin{equation}%
\begin{array}
[c]{c}%
k^{(T)}=k\dfrac{Z}{\sqrt{g}}+k\sqrt{\dfrac{Z^{2}-g}{g}}\cos\psi,\\
k^{(y)}=\dfrac{p}{\sqrt{g}}k+\dfrac{k}{\sqrt{Z^{2}-g}}\left(  p\dfrac{Z}%
{\sqrt{g}}\cos\psi-\dfrac{\mathcal{L}\sqrt{g}}{|T|}\sin\psi\right)  ,\\
k^{(\phi)}=\mathcal{L}\dfrac{k}{|T|}+\dfrac{k}{\sqrt{Z^{2}-g}}\left(
p\sin\psi+\mathcal{L}\dfrac{Z}{|T|}\cos\psi\right)  .
\end{array}
\label{k}%
\end{equation}

\end{itemize}

\bigskip These relations allow us to find the components of the 4-wavevector
in the tetrad frame (\ref{tetr}), given the components in the comoving frame.
Introducing quantities in the frame of reference of an emitter $q_{e}\equiv
k_{y}^{\prime\prime}=k\cos\psi$ and $l_{e}\equiv k_{\varphi}^{\prime\prime
}=k\sin\psi$, we can write the above transformations in the form
\begin{equation}
k^{T}=\sqrt{q_{e}^{2}+l_{e}^{2}}Z+q_{e}\sqrt{Z^{2}-g}\text{,}%
\end{equation}%
\begin{equation}
q=p\sqrt{q_{e}^{2}+l_{e}^{2}}+\dfrac{1}{\sqrt{Z^{2}-g}}\left(  pZq_{e}%
-g\dfrac{\mathcal{L}}{|T|}l_{e}\right)  , \label{q_expr}%
\end{equation}%
\begin{equation}
l=\mathcal{L}\sqrt{q_{e}^{2}+l_{e}^{2}}+\dfrac{1}{\sqrt{Z^{2}-g}}\left(
p|T|l_{e}+\mathcal{L}Zq_{e}\right)  . \label{l_expr}%
\end{equation}

In partial cases $p=0$ or $\mathcal{L}=0$ these expressions simplify.

\begin{itemize}
\item If $\mathcal{L}=0$:
\begin{gather}
q=p\sqrt{q_{e}^{2}+l_{e}^{2}}+\sigma(p)q_{e}\sqrt{p^{2}+g},\\
l=\sigma(p)|T|l_{e}.
\end{gather}

\item If $p=0$:
\begin{gather}
q=-\sqrt{g}\sigma(\mathcal{L})l_{e},\\
l=\mathcal{L}\sqrt{q_{e}^{2}+l_{e}^{2}}+|T|\sigma(\mathcal{L})q_{e}%
\sqrt{\dfrac{\mathcal{L}^{2}}{T^{2}}+1}.
\end{gather}

\end{itemize}

\qquad Hereafter, $\sigma(f)$ means the sign of quantity $f$.

\section{Behavior of parameters of photon\label{secph} \label{par}}

\qquad In the previous section, we described how quantities $q$ and $l$ are
related to the components of the 4-wavevector in the frame comoving with an
emitter. These parameters represent the energy and angular momentum of a
photon and are conserved along its trajectory. In this section, we analyze the
relation between these quantities and the parameters of the emitter. We start
with an analysis of parameters $q$ and $l$ if they were emitted near horizon.
To this end, we derive the following relations from the exact expressions
(\ref{q_expr}-\ref{l_expr}). They will be compiled in Table \ref{tab}.

\begin{table}[ptb]
\centering
\begin{tabular}
[c]{|l||c|c|c|c|c|}\hline
Condition &  & $q$ &  & $l$ & \\\hline
$p \neq0$, $\mathcal{L} \neq0$ & Horizon & $pk(1 +\cos\psi)$ & F &
$\mathcal{L}k(1 +\cos\psi) +\sigma(p)k r_{+} \sin\psi$ & F\par\\\hline
& Singularity & If $\psi\neq0 ,\pi: -\sigma(\mathcal{L})k\sqrt{g}\sin\psi$ &
D & $\mathcal{L}k(1 +\cos\psi)$ & F\\\hline
&  & If $\psi=0 ,\pi:pk(1 +\cos\psi)$ & F & $\mathcal{L}k(1 +\cos\psi)$ &
F\\\hline
$p =0$, $\mathcal{L} \neq0$ & Horizon & $-\sigma(\mathcal{L})k\sqrt{g}\sin
\psi$ & 0 & $\mathcal{L}k(1 +\sqrt{\frac{\mathcal{T}^{2}}{\mathcal{L}^{2}}
+1}\cos\psi)$ & F\\\hline
& Singularity & If $\psi\neq0 ,\pi: -\sigma(\mathcal{L})k\sqrt{g}\sin\psi$ &
D & $\mathcal{L}k(1 +\cos\psi)$ & F\\\hline
&  & If $\psi=0 ,\pi:0$ & 0 & $\mathcal{L}k(1 +\cos\psi)$ & F\\\hline
$p \neq0$, $\mathcal{L} =0$ \par & Horizon & $pk(1 +\cos\psi)$ & F &
$\sigma(p)kr_{+} \sin\psi$ & F \par\\\hline
& Singularity\par & If $\psi\neq\pm\frac{\pi}{2} :k\sigma(p)\sqrt{g}\cos\psi$
& D & $\sigma(p)k r_{+} \sin\psi$ & 0\\\hline
&  & If $\psi= \pm\frac{\pi}{2} :kp$ & F & $\sigma(p)k r_{+} \sin\psi$ &
0\\\hline
$p =0$, $\mathcal{L} =0$ \par & Horizon & 0 & 0 & $\lim_{\mathcal{L}%
\rightarrow0}\lim_{p\rightarrow0}: k r_{+} \cos\psi$ & 0\\\hline
~ & ~ & 0 & 0 & $\lim_{p\rightarrow0}\lim_{\mathcal{L}\rightarrow0}: 0$ &
0\\\hline
& Singularity & 0 & 0 & 0 & 0\\\hline
\end{tabular}
\caption{ Table, showing asymptotic behavior of $q$ and $l$, depending on
parameters of emitter near horizon and singularity. Here, $F$ means "finite",
$D$-"diverges", $0$-"tends to zero". This table allows us to analyze behavior
of $q$ and $l$ only in two limiting cases near horizon and singularity. }%
\label{tab}%
\end{table}

As was shown in \cite{gr20}, the sign of $q$ significantly affects the
behavior of the frequency of radiation absorbed near the horizon and the sign
of $l$ affects the behavior of the frequency of a photon recieved near the
singularity. We would like to stress that, although a photon emitted near the
singularity will be absorbed soon, it can affect the metric significantly if a
corresponding frequency measured by a reciever diverges. Such a case is
possible \cite{gr20}. However, an important question about backreaction of
raidation on the vacuum metric is beyond the scope of our work and needs
separate treatment.

Thus there appears a question if it is possible to emit photons with given
signs of aforementioned parameters. The sign of these quantities can be
altered by emitting the photon at different angles $\psi.$ To analyze the
potential ranges of angle $\psi$ in which quantities $q$ and $l$ can have a
given sign, we first need to solve the equations $q=0$ and $l=0$ for angle
$\psi$.

The solutions to these equations are as follows:
\begin{gather}
q=0\text{:}~\tan\psi=\dfrac{p\sqrt{g}}{|T|}\dfrac{Z\mathcal{L}\sqrt{g}%
\pm|T|(Z^{2}-g)}{(Z^{2}-g)(g-p^{2})-gp^{2}},\label{q0}\\
l=0\text{:}~\tan\psi=-\mathcal{L}|T|\dfrac{Zp\pm(Z^{2}-g)}{(Z^{2}%
-g)(T^{2}-\mathcal{L}^{2})-g\mathcal{L}^{2}}.
\end{gather}

\bigskip%
\begin{figure}
[ptb]
\begin{center}
\includegraphics[
height=3.4739in,
width=6.40679in
]%
{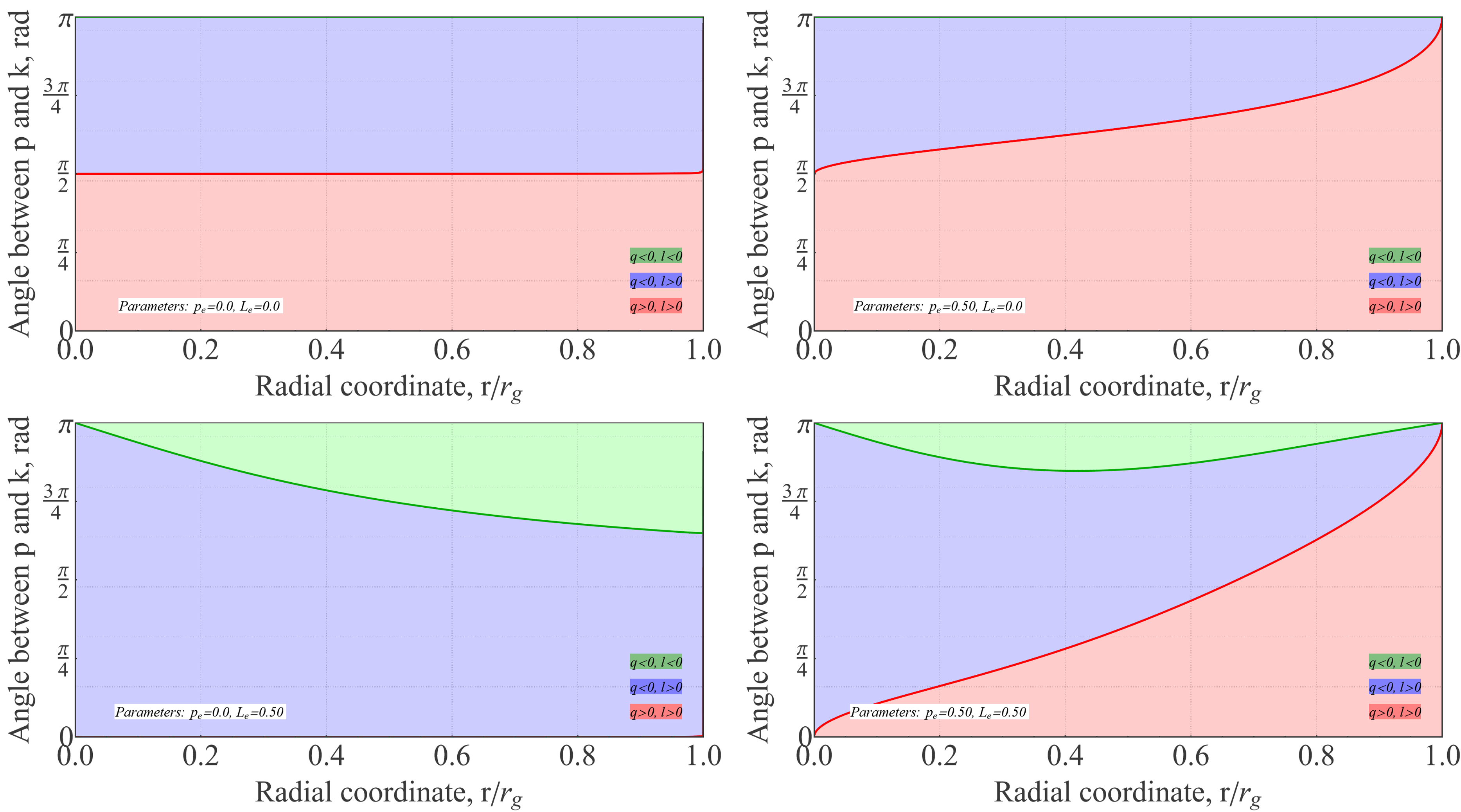}%
\caption{Diagrams, showing in which range of emission angle and emission point
we can get positive or negative values of $q$ or $l.$ Note that for first
diagram we initially took limit $\mathcal{L}\rightarrow0$, and only then
$p\rightarrow0$ (this reservation is important because depending on order of
taking limits we get different results, see Section \ref{par_cas}). If some
color is not presented on a diagram, then it means that such such
configuration is not possible for a given set of parameters.}%
\label{ql_diag}%
\end{center}
\end{figure}

On FIG. \ref{ql_diag} we show how the signs of $q$ and $l$ depend on the
direction of emission and the emission point for various parameters of the
emitter. This analysis enables us to select the direction of radiation (in a
frame comoving with the emitter) based on the position and parameters of the
emitter, such that $q$ and $l$ have specified signs. Here, we will focus on
the most intriguing regions, namely near the horizon and the singularity.

\subsection{Particular cases\label{par_cas}}

$\mathcal{L}=0$, $p\neq0$%

\begin{gather}
q=0\text{:}~\tan\psi=\pm\frac{\sqrt{g}}{p},\\
l=0\text{:}~\psi=0\text{ or }\psi=\pi.
\end{gather}

$\mathcal{L}\neq0$, $p=0$,
\begin{equation}
q=0\text{: }\psi=0\text{ or }\pi,
\end{equation}%
\begin{equation}
l=0\text{: }\tan\psi=\pm\frac{|T|}{\mathcal{L}}.
\end{equation}

$\mathcal{L}=0$, $p=0$%
\begin{equation}
q=0\text{:}~\lim_{p\rightarrow0}\lim_{\mathcal{L}\rightarrow0}\psi=\frac{\pi
}{2}\text{ or}-\frac{\pi}{2},
\end{equation}%
\begin{equation}
q=0\text{:}~\lim_{\mathcal{L}\rightarrow0}\lim_{p\rightarrow0}\psi=0\text{ or
}\pi,
\end{equation}

\begin{equation}
l=0\text{: }~\lim_{p\rightarrow0}\lim_{\mathcal{L}\rightarrow0}\psi=0\text{ or
}\pi,
\end{equation}%
\begin{equation}
l=0\text{:}~\lim_{\mathcal{L}\rightarrow0}\lim_{p\rightarrow0}\psi=\frac{\pi
}{2}\text{ or}-\frac{\pi}{2}.
\end{equation}

\subsection{Horizon}

$\mathcal{L}\neq0$, $p\neq0$,
\begin{equation}
q=0\text{: }\psi=0\text{ or }\psi=\pi,
\end{equation}%
\begin{equation}
l=0\text{: }\tan\psi=-\dfrac{\mathcal{L}|T|}{|T|^{2}-\mathcal{L}^{2}}%
(\sigma(p)\pm1).
\end{equation}
$\mathcal{L}\neq0$, $p=0$
\begin{equation}
q=0\text{: }\psi=0\text{ or }\psi=\pi,
\end{equation}%
\begin{equation}
l=0\text{: }\lim_{p\rightarrow0}\lim_{g\rightarrow0}\tan\psi=\pm
\frac{2\mathcal{L}|T|}{|T|^{2}-\mathcal{L}^{2}}\text{ or }\psi=0\text{, }\pi,
\end{equation}%
\begin{equation}
l=0\text{: }\lim_{g\rightarrow0}\lim_{p\rightarrow0}\tan\psi=\pm\frac
{|T|}{\mathcal{L}}.
\end{equation}

$\mathcal{L}=0$, $p\neq0$%
\begin{equation}
q=0\text{: }\psi=0\text{ or }\psi=\pi,
\end{equation}%
\begin{equation}
l=0\text{: }\psi=0,\pi.
\end{equation}

$p=0$, $\mathcal{L}=0$%

\begin{equation}
q=0\text{: if we take limits in such order: }\lim_{g\rightarrow0}%
\lim_{p\rightarrow0}\lim_{\mathcal{L}\rightarrow0}\text{ then }\psi=\pm
\frac{\pi}{2}.
\end{equation}

\begin{equation}
\text{For all other orders of taking limits we get }\psi=0\text{ or }\psi=\pi,
\end{equation}

\begin{equation}
l=0\text{: if the first taken limit is }\lim_{p\rightarrow0}\text{ then }%
\psi=\pm\frac{\pi}{2}.
\end{equation}

\begin{equation}
\text{For all other orders of taking limits we get }\psi=0\text{ or }\psi=\pi.
\end{equation}

\subsection{Singularity}

$\mathcal{L}\neq0$, $p\neq0$,%
\begin{equation}
q=0\text{: }\psi=0\text{ or }\psi=\pi,
\end{equation}%
\begin{equation}
l=0\text{: }\psi=0\text{ or }\psi=\pi.
\end{equation}

$\mathcal{L}\neq0$, $p=0$
\begin{equation}
q=0\text{: }\psi=0\text{ or }\psi=\pi,
\end{equation}%
\begin{equation}
l=0:\psi=0\text{ or }\psi=\pi.
\end{equation}

$~\mathcal{L}=0$, $p\neq0$%
\begin{equation}
q=0\text{: }\lim_{g\rightarrow\infty}\lim_{\mathcal{L}\rightarrow0}\psi
=\frac{\pi}{2}\text{ or}-\frac{\pi}{2},
\end{equation}%
\begin{equation}
q=0\text{: }\lim_{\mathcal{L}\rightarrow0}\lim_{g\rightarrow\infty}%
\psi=0\text{ or }\pi,
\end{equation}

\begin{equation}
l=0\text{: }\psi=0\text{ or }\psi=\pi.
\end{equation}

$~\mathcal{L}=0$, $p=0$%

\begin{equation}
q=0\text{: if the first taken limit is }\lim_{\mathcal{L}\rightarrow0}\text{
then }\psi=\pm\frac{\pi}{2}.
\end{equation}

\begin{equation}
\text{For all other orders of taking limits we get }\psi=0\text{ or }\psi=\pi.
\end{equation}

\begin{equation}
q=0\text{: if we take limits in such order: }\lim_{g\rightarrow\infty}%
\lim_{\mathcal{L}\rightarrow0}\lim_{p\rightarrow0}\text{ then }\psi=\pm
\frac{\pi}{2}.
\end{equation}

\begin{equation}
\text{For all other orders of taking limits we get }\psi=0\text{ or }\psi=\pi
\end{equation}

All these results correlate with the behavior on (FIG. \ref{ql_diag}).

\section{Frequency of a registered photon \label{fr}}

\qquad The frequency measured by an observer with the four-velocity $u^{\mu} $
is given by $\omega=-u_{\mu}k^{\mu}$. If both the photon and the observer are
moving in the same plane,%
\begin{equation}
\omega=\dfrac{zZ-pq}{g}-\dfrac{l\mathcal{L}}{T^{2}}\text{.} \label{pl}%
\end{equation}
This expression can be derived from equations (\ref{u}) and (\ref{km}), and it
is consistent with eq. (26) of \cite{gr20}.

Let us consider a scenario where a photon propagates at an angle $\alpha$ with
respect to the plane $\theta=\frac{\pi}{2}$ in which an observer is moving. In
this case
\begin{equation}
k^{\mu}=\left(  z,\dfrac{q}{g},\dfrac{l}{T^{2}}\cos\alpha,\dfrac{l}{T^{2}}%
\sin\alpha\right)  ,
\end{equation}
and the formula for the frequency becomes
\begin{equation}
\omega=\dfrac{zZ-pq}{g}-\dfrac{l\mathcal{L}}{T^{2}}\cos\alpha. \label{np}%
\end{equation}

For the motion of a photon and an observer in the same plane, the analytical
results based on eq. (\ref{pl}) were discussed in \cite{gr20}. If the motion
occurs in different planes, we concentrate on the novel features that arise
due to a nonzero $\alpha$ in (\ref{np}). The behavior of the frequency near
the horizon for $\alpha=0$ \cite{gr20} and $\alpha\neq0$ is similar. As far as
the behavior near the singularity is concerned, a qualitatively new case
absent in \cite{gr20} appears now. Namely, if $l\mathcal{L}>0$, for $\alpha=0$
the frequency remains finite \cite{gr20}. However, this is not so for
$\alpha\neq0$.

For massive particles, the effect of nonzero $\alpha$ on particle collisions
near the singularity was considered in Sec. 7 of Ref. \cite{3}.

\section{\bigskip View of night sky \label{sky}}

Now, as we established main relations between parameters of an observer,
emitter and photon, let us apply them to an interesting issue - view of a
night sky, seen by an observer. To this end, we assume that sky is full of
stars which are infinitely far from a black hole. Each star emits photons in
different directions. The photons that passed though the horizon and were
trapped by black hole can be observed by an observer falling to singularity.
To analyze, how these photons move, let us find the angle between the angular
and radial components of the four- momentum of photon:
\begin{equation}
\tan\beta=\frac{k^{\varphi}}{k^{y}}=\frac{l}{q}\frac{\sqrt{g}}{|T|}.
\label{tan}%
\end{equation}

As one can see, near the singularity ($T\rightarrow0$, $g\rightarrow\infty$)
we get $\tan\beta\rightarrow\infty$. This means that near the singularity the
angular component of the wave vector of a photon is much higher than the
$y-$component and three-wave vectors of all photons are practically
perpendicular to $y-$axis. Thus any non-orbiting observer will see the whole
night sky "squeezed" to a ring perpendicular to direction of his/her motion.
However, if an observer has an angular momentum, near the singularity his
3-velocity will also be perpendicular to a $y-$axis. As a result, his/her
three-velocity will lie in a plane of motion of photons, thus "squeezed" ring
will turn along direction of motion.

Consideration above allow us to predict a view of night sky only near the
singularity. To obtain it everywhere under the horizon, we should use the
exact formula for light aberration. The case of a non-orbiting observer was
already analysed in \cite{tw} (see page B-22). We would like to obtain the
same equation for an orbiting observer. We will do it in two steps. First of
all, we will find the angles between direction of motion of a given photon and
an observer in the tetrad frame (\ref{tetr}).\ Afterwards, we will conduct a
boost in the direction of motion. To do this, let us introduce 2 angles
$\gamma$ and $\delta$ describing direction of motion of photon$.$ The first
quantity describes an angle between a spacial part of the wave vector and
coordinate $y$, while $\delta$ describes an angle between the equatorial plane
(in which an observer moves) and the plane of motion of photon. The angle
$\gamma$ was already found in \cite{tw} and is given by an eq. [38] on page
B-23 where it is denoted $\psi_{shell}$.

Thus, the corresponding spacial unit vector describing the direction of a
photon's propagation is given in the tetrad frame (\ref{tetr}) by the expression%

\begin{equation}
h^{(i)}=\frac{k^{(i)}}{|k|}=(\cos\gamma,\sin\gamma\cos\delta,\sin\gamma
\sin\delta). \label{h_exp}%
\end{equation}

The angle between the direction of propagation of a photon and an observer can
be found using the standard definition (note that only in this section we will
denote this angle as $\psi$ that does not coincide with angle $\psi$
introduced in (\ref{k_fr})):%

\begin{equation}
\cos\psi=\frac{v_{(i)}h^{(i)}}{|v||h|},
\end{equation}

$v_{i}$ is a 3-velocity of an observer in the tetrad frame (\ref{tetr}):%
\begin{equation}
v_{(i)}=-\frac{e_{(i)\mu}u^{\mu}}{e_{(0)\mu}u^{\mu}}\text{.}%
\end{equation}
\qquad\qquad Taking into account (\ref{tetr}) and (\ref{u}), we have%

\begin{equation}
v_{(i)}=\left(  \frac{p}{Z},\sqrt{g}\frac{\mathcal{L}}{TZ},0\right)
,|v|=\frac{\sqrt{p^{2}+g\mathcal{L}^{2}/T^{2}}}{\sqrt{p^{2}+g(1+\mathcal{L}%
^{2}/T^{2})}}. \label{3vel}%
\end{equation}

Eventually, we have:\newline%

\begin{equation}
\cos\psi=\left(  p\cos\gamma+\sqrt{g}\frac{\mathcal{L}}{T}\sin\gamma\cos
\delta\right)  \frac{1}{\sqrt{p^{2}+g\mathcal{L}^{2}/T^{2}}}. \label{ps_shell}%
\end{equation}

Note, that if there is no rotation ($\mathcal{L}=0$), then we get $\cos
\psi=\cos\gamma$ (as was noted in analysis of equation [38] on page B-23 in
\cite{tw}).

However, in the case of a non-zero angular momentum, the angle $\psi$ is not
sufficient to describe the position of a star on a celestial sphere of an
observer. To this end, we introduce also an azimuthal angle on celestial
sphere. To find it, let us decompose vector $h^{(i)}$ into two components:
parallel to $v_{(i)}$ and $\widetilde{h}^{(i)}$ that is perpendicular to it,%

\begin{equation}
\widetilde{h}^{(i)}=h^{(i)}-\frac{v^{(i)}}{|v|^{2}}(h^{(k)}v_{(k)}).
\label{htild}%
\end{equation}
Then, an azimuthal angle is defined as the angle between $\widetilde{h}^{(i)}
$ and the vector $n^{(i)}=(0,0,1)$ (where $n^{(i)}$ is a normal vector,
perpendicular to the observer's plane of motion).

This allows us to obtain the azimuthal angle%

\begin{equation}
\cos\varphi=\frac{\widetilde{h}_{(i)}n^{(i)}}{|\widetilde{h}||n|}.
\label{phi_shell}%
\end{equation}

\bigskip Note that the expressions for $\psi$ and $\varphi$ describe the
angles in the tetrad frame (\ref{tetr}). To obtain the picture seen by an
observer, we have to perform boost in the direction of his motion. Note that
angle $\varphi$ will not change during boost, because it is defined through
the scalar product of two vectors, perpendicular to the direction of boots. As
we perform the boost along the observer's velocity, it does not change vectors
perpendicular to the velocity. As the angle $\psi$\ is defined to be an angle
between the velocity and the direction to the star (in the tetrad frame
(\ref{tetr})) we can use a formula from special relativity for change of
angles (which coincides with eq. [34] on page B-22 of \cite{tw}):%

\begin{equation}
\cos\psi^{\prime}=\frac{\cos\psi+|v|}{1+|v|\cos\psi}, \label{psi_pr_shell}%
\end{equation}
(where $|v|$ is given by (\ref{3vel})). Direct derivation of eq.
(\ref{psi_pr_shell}) is carried out in Appendix. Using eq. (\ref{ps_shell}),
one obtains final expressions for the angles $\psi^{\prime}$ and $\varphi$,
under which an observer sees a given star. Because of complexity of
corresponding expressions, we will present the results of numerical
computations only.%

\begin{figure}
[ptb]
\begin{center}
\includegraphics[
height=6.1031in,
width=6.1022in
]%
{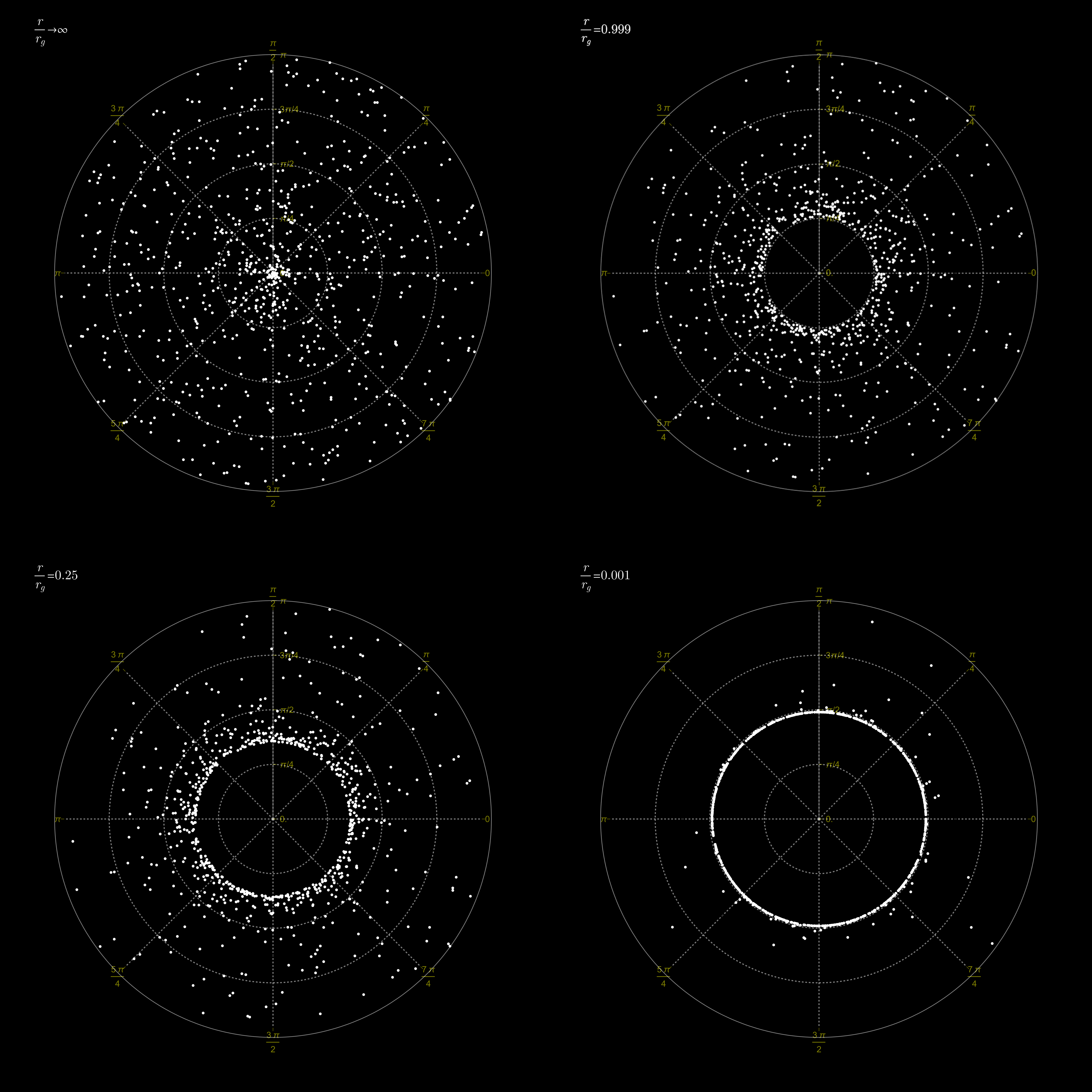}%
\caption{Picture showing how observer will see a night sky as he/she
approaches singularity. On each scheme radial coordinate corresponds to zenith
angle on celestial sphere (angle $\psi^{\prime}$), angular-azimuthal (angle
$\varphi$). Parameters of an observer: $p=-1.05,\mathcal{L}=0$}%
\label{non_rot_sky}%
\end{center}
\end{figure}

\begin{figure}
[ptb]
\begin{center}
\includegraphics[
height=6.1031in,
width=6.1022in
]%
{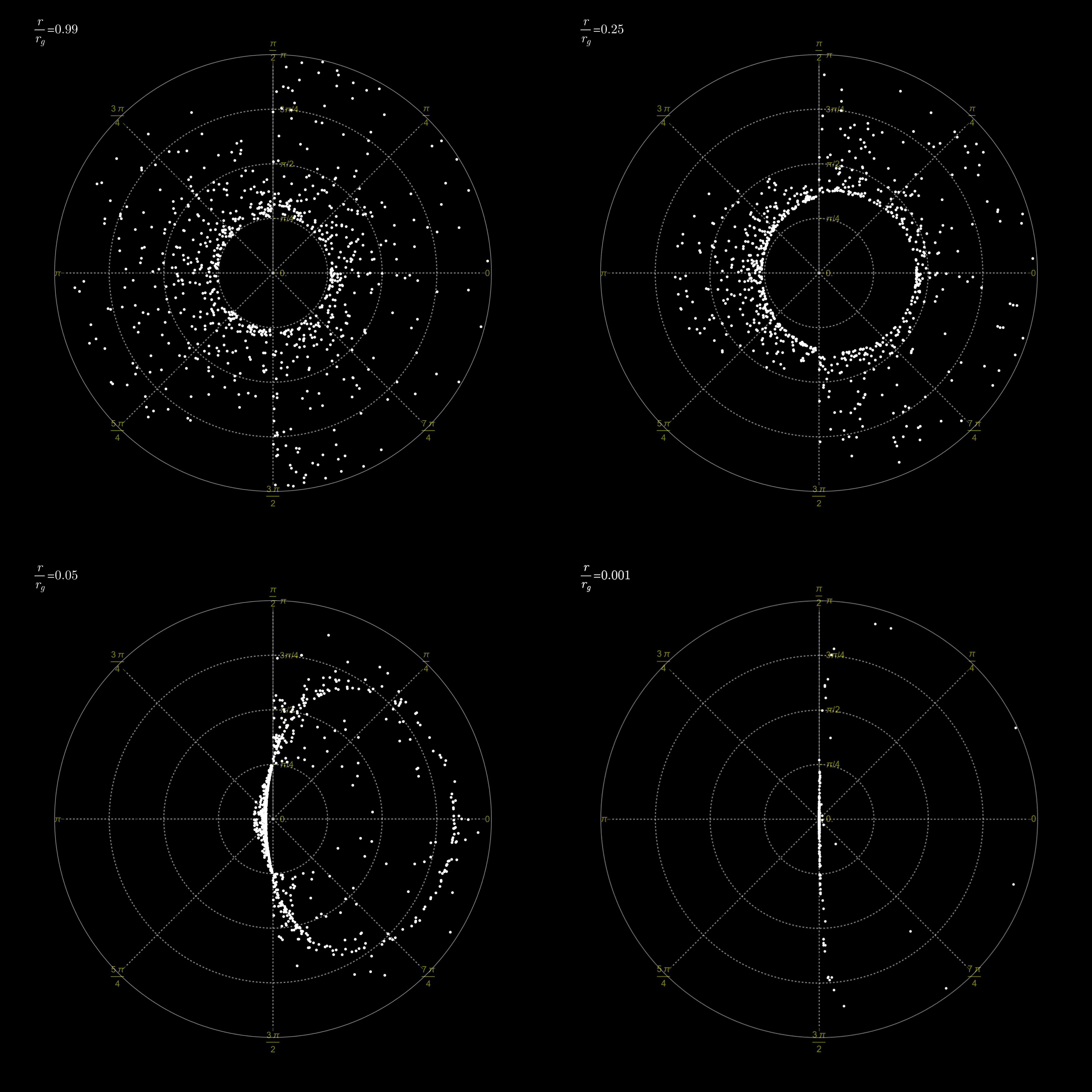}%
\caption{Picture showing how observer will see a night sky as he/she
approaches singularity. On each scheme radial coordinate corresponds to zenith
angle on celestial sphere (angle $\psi^{\prime}$), angular-azimuthal (angle
$\varphi$). View of a night sky, seen from infinity is the same as shown on
Fig (\ref{non_rot_sky}). Parameters of an observer: $p=-1.05,\mathcal{L}%
=0.05$}%
\label{rot_sky}%
\end{center}
\end{figure}

For our numerical computations we generate randomly distributed points on a
sky and analyze how this picture changes as an observer approaches the
singularity. To this end, from each star and for a given position of an
observer, we find numerically an impact parameter $b=l/q$ which a photon
emitted by this star has to have if it reaches an observer at a given point
(photons with different impact parameters can be thought of as photons,
emitted in different directions). This was done by solving equations of motion
for a photon and an observer and by finding such impact parameter that for a
given coordinate $y$ positions of a photon and an observer coincide. Then, we
use eqs. (\ref{phi_shell}) and (\ref{psi_pr_shell}) to obtain angles under
which an observer will see a given star on his/her celestial sphere.

The corresponding results are presented on a Fig. \ref{non_rot_sky} and Fig.
\ref{rot_sky}. On Fig. \ref{non_rot_sky} we depict a view of a night sky seen
by a non-orbiting observer as he/she approaches singularity. Here on each
polar plot the "radial" coordinate corresponds to the zenith angle on
celestial sphere, while "angular" corresponds to the azimuthal one. As we can
see, near the singularity the whole night sky indeed "squeezes" to a ring
perpendicular to the direction of motion in agreement with what we discussed
earlier. One reservation is in order. If an observer moves strictly radially,
he sees radial rays without aberration, so in the center of \ref{non_rot_sky}
a lumineous point (not shown) is present.

\begin{figure}
[ptb]
\begin{center}
\includegraphics[
height=4.9026in,
width=4.9026in
]%
{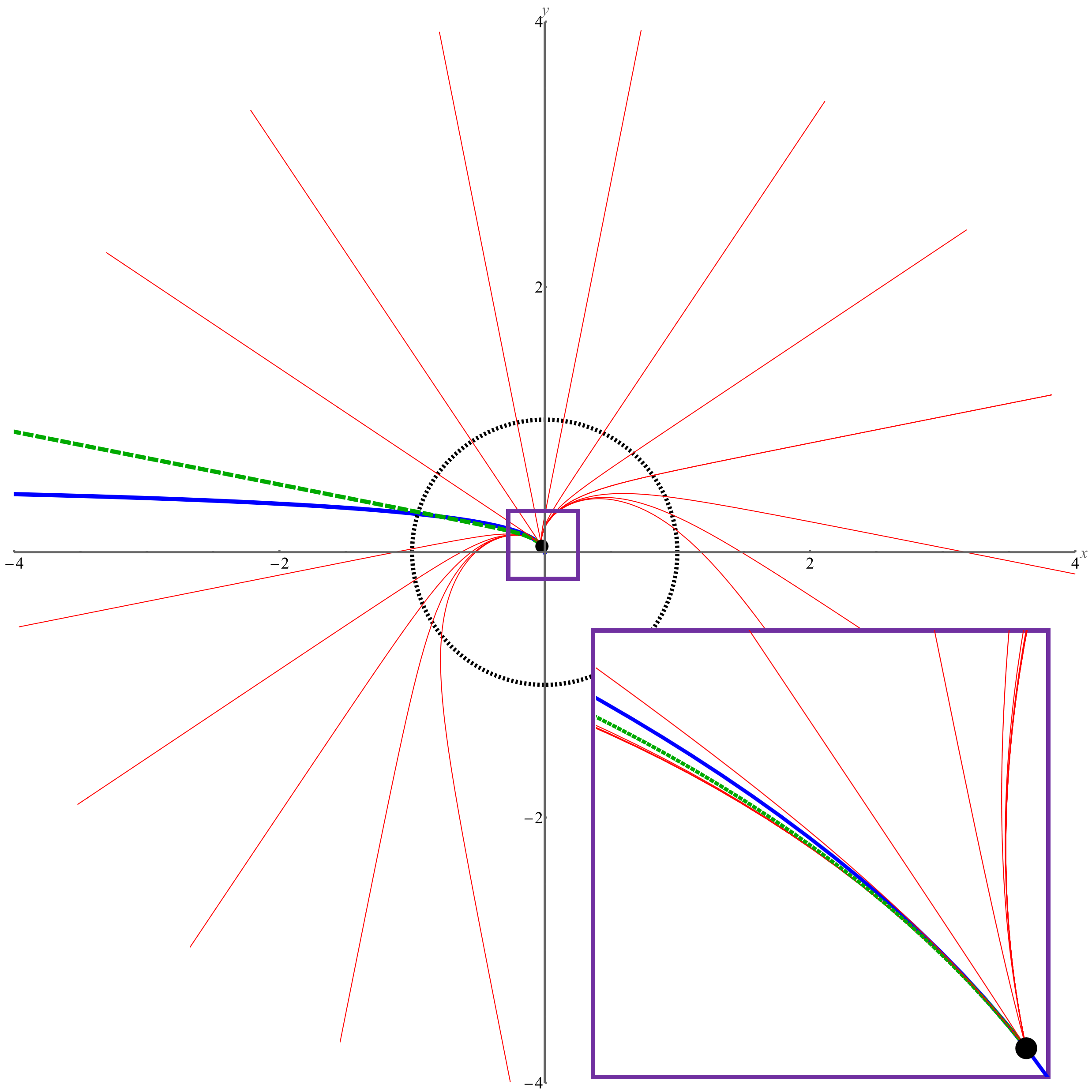}%
\caption{Picture showing trajectory (blue curve) of an orbiting observer whose
current position is presented as a black dot and trajectories of photons (red
and green dased curves). Green dased trajectory represents tajectory of a
photon from a star, which was seen at a left hand side, but approaching
singularity this star is seen at right hand side (left and right hand sides
are defined with respect to line, tangent to observer's trajectory). Dotted
line represents a horizon. Parameters of the observer: $p=-1.05,\mathcal{L}%
=0.05$}%
\label{traj}%
\end{center}
\end{figure}

If an observer has a non-zero angular momentum, we obtain results depicted on
a Fig. \ref{rot_sky}. We see that indeed near the singularity the whole night
sky "squeezes" to a ring lying in a plane generated from the direction of
motion and vertical axis (perpendicular to the motion plane). Also several
interesting observations may be made in this case. First of all, we see that
for both a non-orbing observer and for the orbiting one there exists the inner
"blind zone" around a singularity that has a form of disk in the case of
non-orbiting observer ("blind zones" are regions in which an observer does not
see any radiation). Also, in the case of an orbiting observer there exists
another "blind zone", additional to that aroung the singularity, placed at the
observer's left hand side. The reason of such an effect is rotation of an
observer that leads to the change of a side on which observer sees a given
star (for explanation see Fig. \ref{traj}. Also note that corresponding
trajectories were drawn as polar plots in a parametric space with parameters
$r$ and $\varphi$. This reservation is needed because under the horizon $r$
coordinate is no longer spacelike). Note that this phenomenon was not analysed
in previous works (see \cite{ham},\cite{rizu}) and is of special interest.
Also note that if observer will have another sign of angular momentum, then
aforementioned "blind zone" will appear at his/her right hand side.%

\section{Surface of a collapsing star seen by an observer \label{star}}

\begin{center}%
\begin{figure}
[ptb]
\begin{center}
\includegraphics[
height=3.3598in,
width=6.0502in
]%
{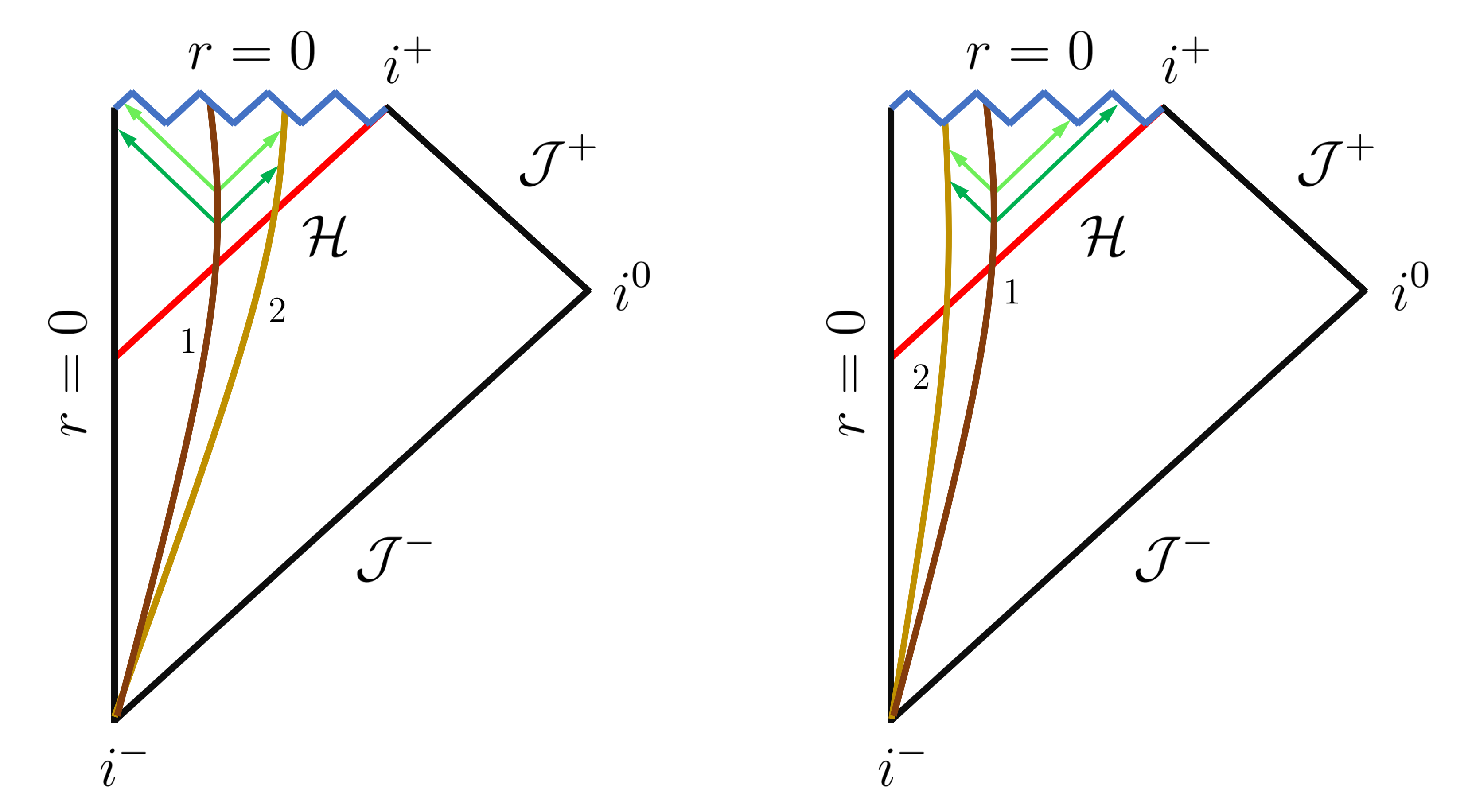}%
\caption{Penrose diagram for a collapsing star. 1 denotes star, 2-external
observer. Arrows represent two photons, which star emits in an opposite
directions}%
\label{pen_diag}%
\end{center}
\end{figure}

\end{center}

\qquad The previous analysis was a preparation for the investigation of a more
interesting physical case - the collapse of a star and radiation from it. We
consider the surface of a collapsing star (which we will denote as number 1 or
with index e) and an observer (2 or index o). After crossing the horizon, the
star emits photons in two opposite directions in a space-time diagram that
will inevitably fall towards the singularity (FIG. \ref{pen_diag}). The
observer (2) may cross the horizon before or after the star. Let us suppose
that the emitter radiates photons in all possible spatial directions during
its fall. An observer at a given position can only observe one of the emitted
photons represented on FIG. \ref{pen_diag}. Our purpose is to relate the
position of an observer with the position of an emitter and the direction of
emission. This will allow us to analyze the frequency of radiation when an
observer falls towards the singularity. With understanding of qualitative
features, let us formulate the governing equations that will allow us to
restore the full picture of the frequency behavior.

First of all, let us investigate the trajectory of a collapsing star. As the
four-velocity of star is given by (\ref{u}), $\dfrac{dT_{e}}{dy_{e}}%
=\dfrac{u^{T}}{u^{y}}=g\dfrac{Z_{e}}{p_{e}}$, whence
\begin{equation}
y_{e}=\int\nolimits_{T_{H}}^{T_{e}}\dfrac{p_{e}}{g}\dfrac{dT}{Z_{e}},
\label{y_e}%
\end{equation}
where $T_{H}=-r_{+}$ and integration is provided from the horizon to a current
position of an emitter. The same equation holds for an observer:
\begin{equation}
y_{o}=\int\nolimits_{T_{H}}^{T_{o}}\dfrac{p_{o}}{g}\dfrac{dT}{Z_{o}}.
\label{y_o}%
\end{equation}
The proper time $\tau_{e}$ of an observer elapsed between the moments $T_{H}$
and $T_{o}$ is equal to
\begin{equation}
\tau_{e}=\int\nolimits_{T_{H}}^{T_{o}}\dfrac{dT}{\sqrt{p_{e}^{2}+g(T)\left(
1+\dfrac{\mathcal{L}_{e}}{T^{2}}\right)  }}\text{.} \label{tau}%
\end{equation}

For an emitter, the similar formula holds with the index "$o$" replaced by
"$e$"$\text{.}$

For a photon that is emitted at $T_{e}$ and received at $T_{ph}$ we have
\begin{equation}
y_{ph}=\int\nolimits_{T_{e}}^{T_{ph}}\dfrac{q}{g}\dfrac{dT}{z}\text{.}
\label{y_ph}%
\end{equation}

In a similar way, for the angular coordinates using (\ref{u}) we can write
$\frac{d\varphi}{dT}=\frac{u^{\varphi}}{u^{T}}.$ We assume, for definiteness,
that an observer crosses the horizon with $\varphi_{o}=0$ and an emitter so
does with $\left(  \varphi_{e}\right)  _{H}$. Then, it follows that%
\begin{equation}
\varphi_{o}=%
{\displaystyle\int\limits_{T_{H}}^{T_{o}}}
\frac{\mathcal{L}_{o}}{T^{2}Z_{o}}dT, \label{phio}%
\end{equation}

\begin{equation}
\varphi_{e}=\left(  \varphi_{e}\right)  _{H}+\int_{T_{H}}^{T_{e}}\frac
{L_{e}dT}{T^{2}Z_{e}}, \label{phie}%
\end{equation}

\begin{equation}
\varphi_{ph}=\left(  \varphi_{ph}\right)  _{H}+%
{\displaystyle\int\limits_{T_{H}}^{T_{ph}}}
\frac{l}{T^{2}z}dT. \label{phip}%
\end{equation}

Angular coordinates remain regular on the horizon, corresponding integrals
(\ref{phio}) - (\ref{phip}) converge. However, the coordinates $(T,y)$ become
singular near the event horizon, integrals (\ref{y_e}), (\ref{y_o}) and
(\ref{y_ph}) diverge there. To avoid these divergences, we resort to the
Kruskal coordinates.

\section{Description of radiation and emission in terms of Kruskal coordinates
\label{kr}}

For the region under the horizon $r<r_{+}\,$, the metric has the form%
\begin{equation}
ds^{2}=-\dfrac{dT^{2}}{g}+gdy^{2}+T^{2}d\Omega^{2}.
\end{equation}
Lets us make transformation%
\begin{equation}
U=\exp\left(  -\kappa u\right)  \text{,} \label{ut}%
\end{equation}%
\begin{equation}
V=\exp\left(  \kappa v\right)  \text{,} \label{vt}%
\end{equation}
where
\begin{equation}
u=y-T^{\ast}\text{, }v=y+T^{\ast}, \label{yTU}%
\end{equation}
$\kappa=(2r_{+})^{-1}$ being a surface gravity and $T^{\ast}$ is given by the formula%

\begin{equation}
T^{\ast}=\int^{T}\frac{dT}{g}\text{.} \label{tg}%
\end{equation}
Now let us obtain equations of motion in terms of Kruskal coordinates. To this
end, we use the relations $\frac{dU}{d\tau}=\frac{dU}{du}\frac{du}{d\tau
}=\frac{dU}{du}\left(  \frac{dy}{d\tau}-\frac{dT^{\ast}}{d\tau}\right)  $.

Using expressions for $y$ (\ref{y_e}), $\tau$ (\ref{tau}) and $T^{\ast}$
(\ref{tg}) we get:%

\begin{equation}
\frac{dU}{d\tau}=\frac{p-Z}{g}\frac{dU}{du}\text{.} \label{mu}%
\end{equation}

In a simiular way, one can get%

\begin{equation}
\frac{dV}{d\tau}=\frac{p+Z}{g}\frac{dV}{dv}\text{,} \label{mV}%
\end{equation}

\bigskip where $Z$ is given by (\ref{Z_expr}).

They can be also rewritten as
\begin{equation}
\dot{U}=\frac{(p-Z)}{FV\kappa}\text{,} \label{uu}%
\end{equation}%
\begin{equation}
\dot{V}=-\frac{(p+Z)}{UF\kappa}. \label{vv}%
\end{equation}
with%
\begin{equation}
F=-\frac{g}{UV\kappa^{2}}\text{.} \label{Fg}%
\end{equation}
The metric takes the form%

\begin{equation}
ds^{2}=-FdUdV+r^{2}d\omega^{2}, \label{mF}%
\end{equation}
where $\kappa$ is the surface gravity, $F\neq0$ is finite on the horizon.

Similar equations are valid for photons:%
\begin{equation}
\dot{U}=\frac{(q-z)}{FV\kappa}\text{,}%
\end{equation}%
\begin{equation}
\dot{V}=-\frac{(q+z)}{UF\kappa}.
\end{equation}

Here, dot means derivative with respect to the affine parameter along the
light-like geodesic.

The photon's wave vector in terms of Kruskal coordinates reads%
\begin{equation}
k^{U}=k^{\mu}\frac{dU}{dx^{\mu}}=\frac{\kappa U}{g}(z-q)=\frac{q-z}{F\kappa
V}\text{,}%
\end{equation}

\begin{equation}
k^{V}=k^{\mu}\frac{dV}{dx^{\mu}}=\frac{\kappa V}{g}(z+q)=-\frac{z+q}{F\kappa
U}\text{,}%
\end{equation}%
\begin{equation}
k_{U}=g_{UV}k^{V}=\frac{z+q}{2\kappa U},
\end{equation}%
\begin{equation}
k_{V}=g_{UV}k^{U}=\frac{z-q}{2\kappa V}.
\end{equation}
Remembering that $k_{\phi}=l$ and using (\ref{u}), for the frequency
$\omega=-k_{\mu}u^{\mu}$ we obtain%
\begin{equation}
\omega=\frac{zZ-pq}{\kappa^{2}FUV}-\dfrac{l\mathcal{L}}{T^{2}}. \label{omkr}%
\end{equation}

It follows from (\ref{Fg}) that it coincides with eq. (\ref{pl}) and could be
obtained directly from it using (\ref{ut}) - (\ref{tg}).

\section{Conditions of absorption \label{cond}}

Let a photon be emitted in point E and absorbed by the observer in point A.
For this event to occur, the corresponding cooridnates should coincide.

\subsection{Angular coordinates}

The condition
\begin{equation}
\varphi_{ph}(A)=\varphi_{o}(A), \label{phi_cond}%
\end{equation}
must hold. Here, according to (\ref{phip}),%
\begin{equation}
\varphi_{ph}=\varphi_{e}(E)+\int_{T_{e}}^{T_{ph}}\frac{ldT}{T^{2}z}\text{.}%
\end{equation}

Using (\ref{phi_cond}), we have%
\begin{equation}
\varphi_{e}(E)+\int_{T_{e}}^{T_{A}}\frac{ldT}{T^{2}z}=\varphi_{o}(T_{A}).
\end{equation}

It is equivalent to%
\begin{equation}
\left(  \varphi_{e}\right)  _{H}+\int_{T_{H}}^{T_{e}}\frac{L_{e}dT}{T^{2}%
Z_{e}}+\int_{T_{e}}^{T_{A}}\frac{ldT}{T^{2}z}=\int_{T_{H}}^{T_{A}}\frac
{L_{o}dT}{T^{2}Z_{o}}\text{.}%
\end{equation}

\subsection{Kruskal coordinates}

Along the trajectory of a massive particle,%
\begin{equation}
U=U(V,L,\varepsilon,m;U_{0},V_{0})\text{,}%
\end{equation}
where%
\begin{equation}
U_{0}=U(V=V_{0})\text{.}%
\end{equation}

Then, if a photon is emitted in the point $(U_{e},V_{e})$, in a similar way%

\begin{equation}
U_{ph}=U(V,l,\omega_{0},0,U_{e},V_{e})\text{,}%
\end{equation}
where%
\begin{equation}
U_{e}=U(V_{e},L_{e},\varepsilon_{e},m_{e};U_{eH},V_{eH}).
\end{equation}

Here, it is assumed that it crossed the horizon at $(U_{H},V_{H}).$ For the
future horizon, $U_{H}=0$, so trajectories 1 and 2 differ by $V_{1}\equiv
V_{1H}$, $V_{2}\equiv V_{2H}$. If $V_{1}<V_{2}\,$, particle 1 crosses the
horizon first.

We also have%
\begin{equation}
U_{o}=U_{o}(V,L_{o},\varepsilon_{o},m_{o};U_{0H},V_{0H}).
\end{equation}

In the point $(U_{A},V_{A})$ where a photon is absorbed,
\begin{equation}
U_{ph}(A)=U(V_{A},l,\omega_{0},0,U_{e},V_{e})\text{,}%
\end{equation}%
\begin{equation}
U_{o}(A)=U_{o}(V_{A},L_{o},\varepsilon_{o},m_{o};U_{0H},V_{0H}),
\end{equation}
and%
\begin{equation}
U_{ph}(A)=U_{o}(A).
\end{equation}

\section{Results for frequency and discussion \label{res}}

\qquad Because of the complexity of the general integrals, we have performed
the corresponding computations numerically. Due to the complexity of the
problem and the large number of different parameters involved, it is not
obvious how each parameter influences the behavior of the frequency. Numerical
analysis shows that qualitatively, all physically different cases can be
classified by the frequency's behavior near the horizon and singularity. Based
on the theoretical analysis conducted in \cite{gr20}, it can be determined
that the frequency near the horizon may either be finite or diverge, while
near the singularity, it may be finite, diverge, or tend to zero. Therefore,
we have six fundamentally different cases. To denote them, we use the
following notation: if the frequency near the horizon is finite, it will be
denoted as case I, if it diverges, it will be denoted as case II. The behavior
near the singularity will be denoted as subcases: if the frequency is finite,
it will be denoted as 1, if it diverges, it will be denoted as 2, and if it
tends to zero, it will be denoted as 3. For example, the case where the
frequency near the horizon is finite but diverges near the singularity will be
denoted as I.2, etc. In \cite{gr20} it is shown that these different scenarios
depend crucially on the mutual signs of the momentum and angular momentum of a
photon (we summarize these conditions in Table \ref{tab3}).

\begin{table}[ptb]
\centering%
\begin{tabular}
[c]{|c|c|c|c|c|}\hline
& $\omega$ near horizon & Conditions & $\omega$ near singularity &
Conditions\\\hline
I.1 & Finite & $pq>0$ or $p=q=0$ & Finite & $Ll>0$\\\hline
I.2 & Finite & $pq>0$ or $p=q=0$ & Diverges & $Ll<0$ or $L=0$ or $l=0$\\\hline
I.3 & Finite & $pq>0$ or $p=q=0$ & Tends to zero & $L=l=0$\\\hline
II.1 & Diverges & $pq<0$ or $p=0$ or $q=0$ & Finite & $Ll>0$\\\hline
II.2 & Diverges & $pq<0$ or $p=0$ or $q=0$ & Diverges & $Ll<0$ or $L=0$ or
$l=0$\\\hline
II.3 & Diverges & $pq<0$ or $p=0$ or $q=0$ & Tends to zero & $L=l=0$\\\hline
\end{tabular}
\caption{ Table showing the conditions that must be met for each case of
different behavior of the frequency.}%
\label{tab3}%
\end{table}

\begin{figure}
[ptb]
\begin{center}
\includegraphics[
height=3.1938in,
width=6.0277in
]%
{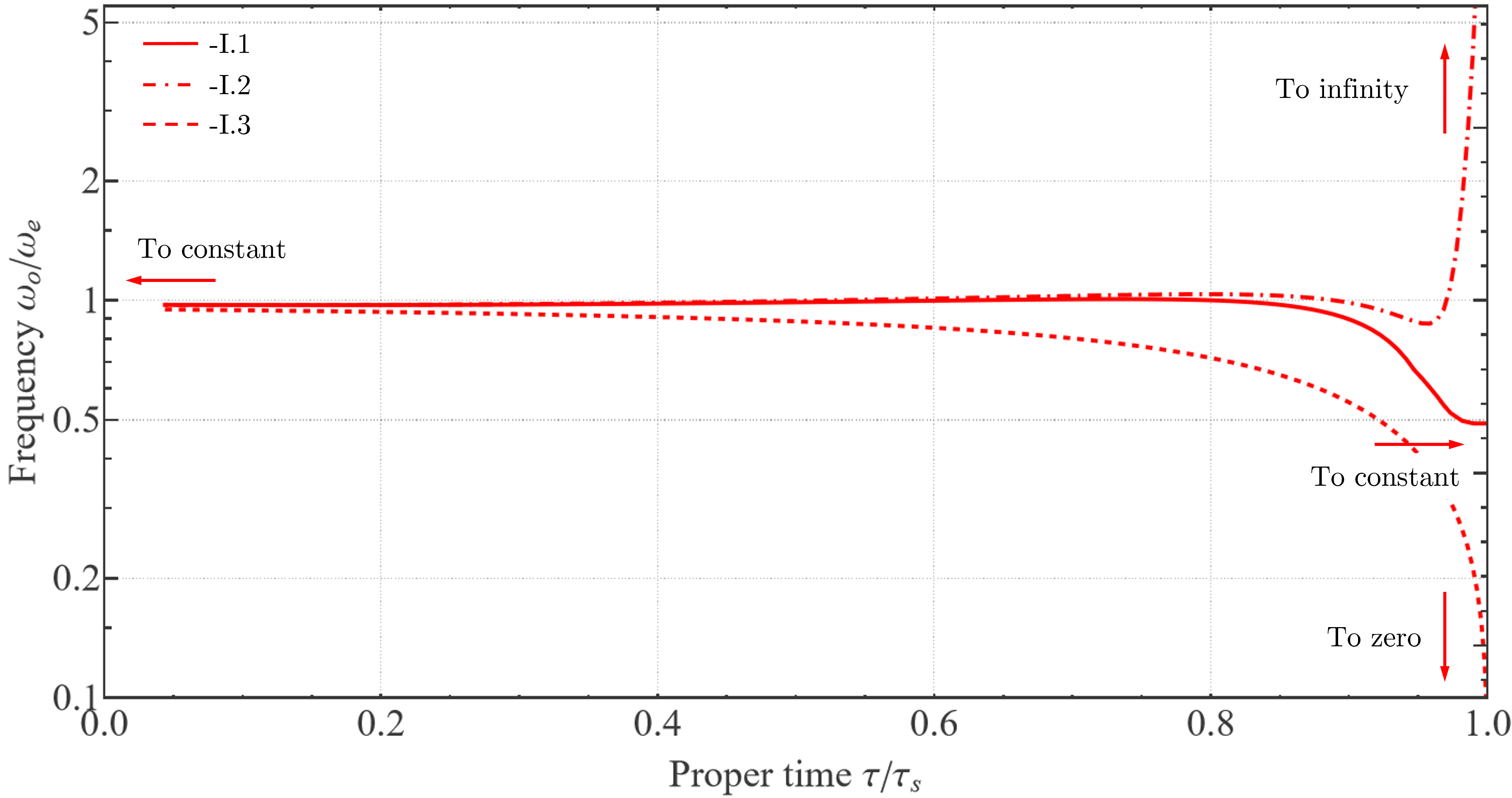}%
\caption{A plot showing the dependence of the frequency obtained by an
observer (normalized to the initial frequency emitted by an emitter) on the
proper time of the observer as it falls towards the singularity. The plot
represents three different scenarios. The parameters for each scenario are as
follows: I.1-$p_{o}=p_{e}=-1.05,\mathcal{L}_{o}=\mathcal{L}_{e}=0,\Delta
\varphi=0,\Delta y=-0.1.$ I.2-$p_{o}=p_{e}=-1.05,\mathcal{L}_{o}%
=\mathcal{L}_{e}=0.1,\Delta\varphi=0,\Delta y=-0.1.$ I.3-$p_{o}=p_{e}%
=-1.05,\mathcal{L}_{o}=-\mathcal{L}_{e}=0.1,\Delta\varphi=0,\Delta y=-0.1.$}%
\label{om_1}%
\end{center}
\end{figure}

\begin{figure}
[ptb]
\begin{center}
\includegraphics[
height=3.1773in,
width=6.0277in
]%
{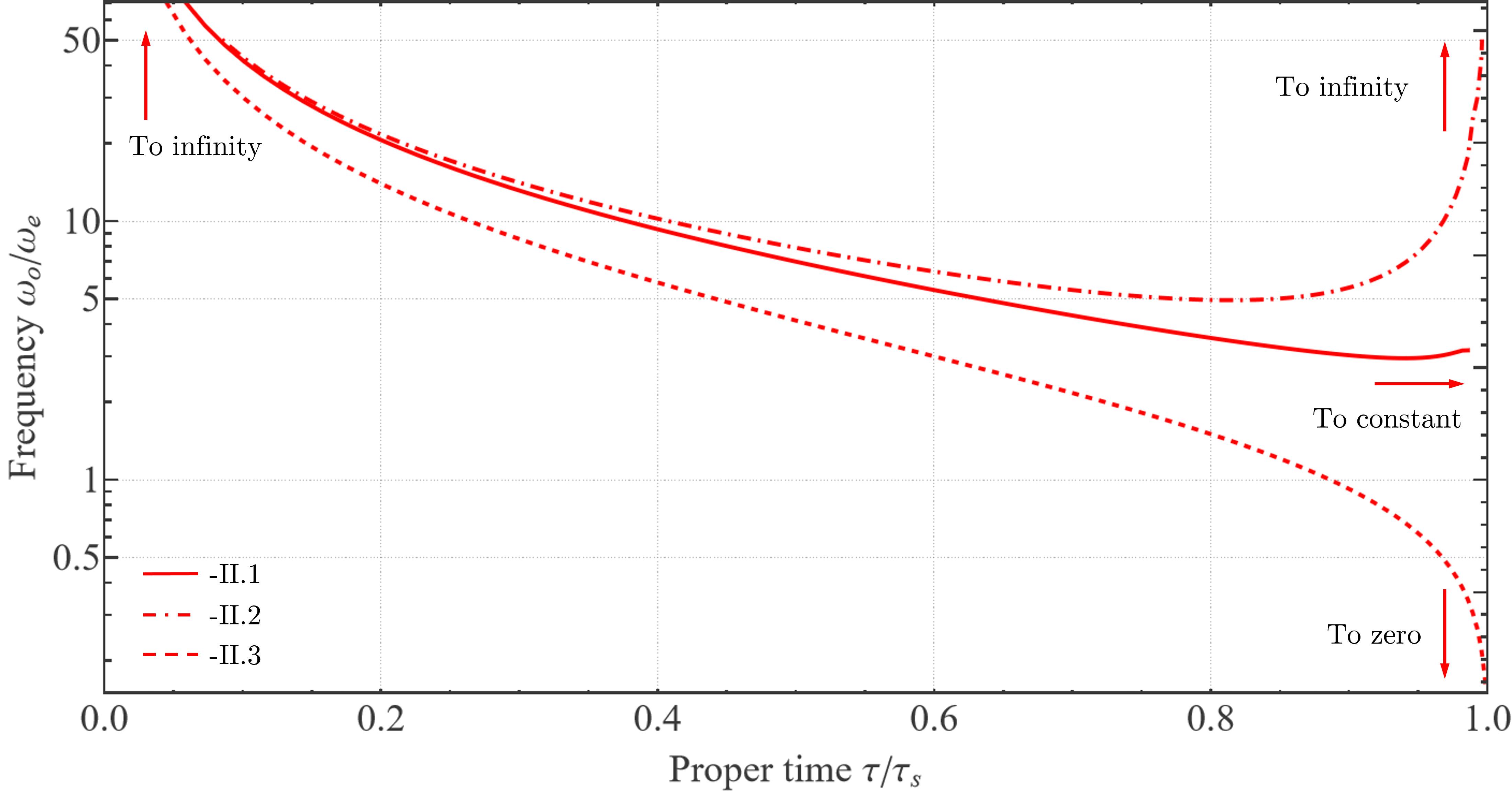}%
\caption{A plot showing the dependence of the frequency obtained by an
observer (normalized to the initial frequency emitted by an emitter) on the
proper time of the observer as it falls towards the singularity. The plot
represents three different scenarios. The parameters for each scenario are as
follows: I.1-$p_{o}=-p_{e}=-1.05,\mathcal{L}_{o}=\mathcal{L}_{e}%
=0,\Delta\varphi=0,\Delta y=-0.1.$ I.2-$p_{o}=-p_{e}=-1.05,\mathcal{L}%
_{o}=\mathcal{L}_{e}=0.1,\Delta\varphi=0,\Delta y=-0.1.$ I.3-$p_{o}%
=-p_{e}=-1.05,\mathcal{L}_{o}=-\mathcal{L}_{e}=0.1,\Delta\varphi=0,\Delta
y=-0.1.$}%
\label{om_2}%
\end{center}
\end{figure}

According to the corresponding results, we must select parameters for the
emitter and observer that satisfy the conditions listed in this Table. The
problem regarding the relationship between the parameters of the emitter and
photon has already been addressed in Section \ref{secph}. With knowledge of
these relationships, we can choose parameters for both the emitter and
observer that allow us to obtain all scenarios of all types included in Table
\ref{tab3}. Typical plots covering all cases in Table \ref{tab3}, are shown on
Fig. \ref{om_1} and Fig. \ref{om_2}. Note that, as we analyzed before, the
behaviour of frequency near the horizon is defined by the signs of momentum of
an emitter $p_{e}$ and that of an observer $p_{o}$. The order in which an
observer and emitter cross the horizon along with requirement of meeting a
photon and observer impose some restrictions. If an observer crossed the
surface of the star before an emitter, then photon has to be emitted in the
direction of emitter's motion that gives the same sign of the photon's
momentum as that of the emitter (see Fig. \ref{ql_diag}). As in this case the
signs of the momentum of an observer coincides with that of a photon, the
frequency is finite. However, if an observer crossed surface of the star after
an emitter, then situation is reversed: a photon has to be emitted in an
opposite direction of emitter's motion (for explanation see Fig.
\ref{pen_diag}) that gives different signs of momenta. The role of other
parameters (such as angular momentum) is more obvious and can be seen from
Table II.

Some reservations are in order. Formally, the case $p>0$ and $q<0$ that leads
to an infinite frequency on the \ horizon is realized when a photon comes from
the "mirror" side of the Schwarzschild metric. However, in a realistic
collapse of the type depicted on Fig. 5 and Fig. 6, such a region does not
arise at all. If we require the frequency to be finite on the horizon, this
enforces the quantity $q$ to be small. See for details Sec. 8.1 in Ref.
\cite{gr20}.

There are also cases when $q=0$ or $p=0$. The first one can be obtained if eq.
(\ref{q0}) is satisfied. The second case is in principle possible but it does
not correspond to a fall of an observer from the outside. See \cite{zero} for details.%

%

\section{Summary} \label{sum}

Thus we analyzed behavior of a frequency and direction under which light is
seen during free fall of a massive observer. We analyzed both cases of
receiving of a photon with some prescribed parameters or the emission by some
other object (for example, surface of a collapsing star). Despite the
complexity of corresponding equations, all cases of behavior of frequency can
be classified according to its the near-horizon and near-singularity behavior.
Different cases may be obtained by manipulating parameters of an emitted
photon. As we showed, this problem can be solved analytically, thus
simplifying overall analysis. Also, we analyzed dynamics of a view of a night
sky and, along with already known results, we discovered an existence of an
additional \textquotedblleft blind zones\textquotedblright\ in a case of an
orbiting observer.

\section{Appendix: proof of aberration formula \label{ap}}

In this section we give direct derivation of aberration formula
(\ref{psi_pr_shell}) under the horizon. To this end, we indroduce the set of
tetrads attached to an observer. We can rewrite expression for 3-velocity
(\ref{3vel}) in a form:%

\begin{equation}
v^{(i)}=|v|(\cos\alpha,\sin\alpha,0),
\end{equation}

where
\begin{equation}
|v|=\sqrt{\frac{p^{2}+g\mathcal{L}^{2}/T^{2}}{p^{2}+g(1+\mathcal{L}^{2}%
/T^{2})}},\tan\alpha=\sqrt{g}\frac{\mathcal{L}}{|T|p}.
\end{equation}

For each point on a trajectory we transform the tetrad attached to an
observer, in such a way:

\begin{itemize}
\item Rotate tetrad in $y\varphi$ plane:%
\begin{equation}
\widetilde{e}_{(0)}=e_{(0)},\widetilde{e}_{(3)}=e_{(3)},
\end{equation}%
\begin{equation}
\widetilde{e}_{(1)}=e_{(1)}\cos\alpha+e_{(2)}\sin\alpha,\widetilde{e}%
_{(2)}=e_{(2)}\cos\alpha-e_{(1)}\sin\alpha.
\end{equation}

\item Perform a boost along $\widetilde{e}_{(1)}:$%
\begin{equation}
\widehat{e}_{(2)}=\widetilde{e}_{(2)},\widehat{e}_{(3)}=\widetilde{e}_{(3),}%
\end{equation}%
\begin{equation}
\widehat{e}_{(0)}=\gamma(\widetilde{e}_{(0)}+|v|\widetilde{e}_{(1)}%
),\widehat{e}_{(1)}=\gamma(\widetilde{e}_{(1)}+|v|\widetilde{e}_{(2)}).
\end{equation}

\end{itemize}

\bigskip After these transformations we obtain:%
\begin{equation}
\widehat{e}_{(0)}=\left(  -\frac{Z}{g},p,\mathcal{L},0\right)  ,
\end{equation}

\begin{equation}
\widehat{e}_{(1)}=\left(  -\frac{\sqrt{p^{2}+g\mathcal{L}^{2}/T^{2}}}{g}%
,\frac{pZ}{\sqrt{p^{2}+g\mathcal{L}^{2}/T^{2}}},\frac{\mathcal{L}Z}%
{\sqrt{p^{2}+g\mathcal{L}^{2}/T^{2}}},0\right)  ,
\end{equation}

\begin{equation}
\widehat{e}_{(2)}=\frac{1}{\sqrt{p^{2}+g\mathcal{L}^{2}/T^{2}}}\left(
0,-g\frac{\mathcal{L}}{|T|},|T|p,0\right)  ,
\end{equation}

\begin{equation}
\widehat{e}_{(3)}=(0,0,0,|T|\sin\theta).
\end{equation}

We use the hat to denote the tetrad attached to an observer. The corresponding
velocity wih respect to it is zero:%

\begin{equation}
\widehat{V}^{(i)}=-\frac{\widehat{e}_{(i)\mu}u^{\mu}}{\widehat{e}_{(0)\mu
}u^{\mu}}=0.
\end{equation}

Now our task is to find an angle between the spacial part of a wavevector and
vector $\widehat{e}_{(1)}$ (which is co-directed with a velocity of an
observer). The wavevector can be written in the form:%

\begin{equation}
k^{\mu}=|k|\left(  -\sqrt{g},\frac{\cos\gamma}{\sqrt{g}},\frac{\sin\gamma
\cos\delta}{|T|},\frac{\sin\gamma\sin\delta}{|T|\sin\theta}\right)  .
\end{equation}
\qquad

\bigskip Components of wavevector, defined in tetrad frame, are given by:%
\begin{equation}
\widehat{k}^{(i)}=-\frac{\widehat{e}_{(i)\mu}k^{\mu}}{\widehat{e}_{(0)\mu
}k^{\mu}}.
\end{equation}

As we already mentioned, we are interested in the angle between wavevector and
direction of motion of an oberver, which is given by:%
\begin{equation}
\widehat{k}^{(1)}=\cos\psi^{\prime}=\frac{\frac{\sqrt{p^{2}+g\mathcal{L}%
^{2}/T^{2}}}{Z}+\frac{1}{\sqrt{p^{2}+g\mathcal{L}^{2}/T^{2}}}\left(
p\cos\gamma+\sqrt{g}\frac{\mathcal{L}}{|T|}\sin\gamma\cos\delta\right)
}{1+\frac{1}{Z}\left(  p\cos\gamma+\sqrt{g}\frac{\mathcal{L}}{|T|}\sin
\gamma\cos\delta\right)  }.
\end{equation}

One can see that this is exactly (\ref{psi_pr_shell}) with $\psi$ given by
(\ref{ps_shell}) and $|v|$ given by (\ref{3vel}).

Data availability

Data sharing not applicable to this article as no datasets were generated or
analysed during the current study.

Ethics declarations

Conflict of interest

The authors have no relevant financial or non-financial interests to disclose.

\end{document}